\documentclass[prb,twocolumn,superscriptaddress,showpacs,preprintnumbers,amsmath,amssymb]{revtex4-1}

\usepackage{graphicx}
\usepackage{dcolumn}
\usepackage{bm}
\usepackage{color,ulem}

\begin{document}


\title{Quantum magnetisms in uniform triangular lattices Li$_2$$A$Mo$_3$O$_8$ ($A$ = In, Sc)}

\author{K.~Iida}\email{k\_iida@cross.or.jp}
\affiliation{Neutron Science and Technology Center, Comprehensive Research Organization for Science and Society (CROSS), Tokai, Ibaraki 319-1106, Japan}

\author{H.~Yoshida}
\affiliation{Department of Physics, Faculty of Science, Hokkaido University, Sapporo, Hokkaido 060-0810, Japan}

\author{H.~Okabe}
\affiliation{Institute of Materials Structure Science, High Energy Accelerator Research Organization (KEK), Tokai, Ibaraki 319-1106, Japan}

\author{N.~Katayama}
\affiliation{Department of Applied Physics, Nagoya University, Nagoya, Aichi 464-8603, Japan}

\author{Y.~Ishii}
\affiliation{Department of Physics, Faculty of Science, Hokkaido University, Sapporo, Hokkaido 060-0810, Japan}

\author{A.~Koda}
\affiliation{Institute of Materials Structure Science, High Energy Accelerator Research Organization (KEK), Tokai, Ibaraki 319-1106, Japan}
\affiliation{Department of Materials Structure Science, Sokendai (The Graduate University for Advanced Studies), Tsukuba, Ibaraki 305-0801, Japan}

\author{Y.~Inamura}
\affiliation{J-PARC Center, Japan Atomic Energy Agency (JAEA), Tokai, Ibaraki 319-1195, Japan}

\author{N.~Murai}
\affiliation{J-PARC Center, Japan Atomic Energy Agency (JAEA), Tokai, Ibaraki 319-1195, Japan}

\author{M.~Ishikado}
\affiliation{Neutron Science and Technology Center, Comprehensive Research Organization for Science and Society (CROSS), Tokai, Ibaraki 319-1106, Japan}

\author{R.~Kadono}
\affiliation{Institute of Materials Structure Science, High Energy Accelerator Research Organization (KEK), Tokai, Ibaraki 319-1106, Japan}
\affiliation{Department of Materials Structure Science, Sokendai (The Graduate University for Advanced Studies), Tsukuba, Ibaraki 305-0801, Japan}

\author{R.~Kajimoto}
\affiliation{J-PARC Center, Japan Atomic Energy Agency (JAEA), Tokai, Ibaraki 319-1195, Japan}

\date{\today}

\begin{abstract}
Molecular based spin-1/2 triangular lattice systems such as LiZn$_2$Mo$_3$O$_8$ have attracted research interest.
Distortions, defects, and intersite disorder are suppressed in such molecular-based magnets, and intrinsic geometrical frustration gives rise to unconventional and unexpected ground states.
Li$_2$$A$Mo$_3$O$_8$ ($A$ = In or Sc) is such a compound where spin-1/2 Mo$_3$O$_{13}$ clusters in place of Mo ions form the uniform triangular lattice.
Their ground states are different according to the $A$ site.
Li$_2$InMo$_3$O$_8$ undergoes conventional $120^\circ$ long-range magnetic order below $T_\text{N}=12$~K whereas isomorphic Li$_2$ScMo$_3$O$_8$ exhibits no long-range magnetic order down to 0.5~K.
Here, we report exotic magnetisms in Li$_2$InMo$_3$O$_8$ and Li$_2$ScMo$_3$O$_8$ investigated by muon spin rotation ($\mu$SR) and inelastic neutron scattering (INS) spectroscopies using polycrystalline samples.
Li$_2$InMo$_3$O$_8$ and Li$_2$ScMo$_3$O$_8$ show completely different behaviors observed in both $\mu$SR and INS measurements, representing their different ground states.
Li$_2$InMo$_3$O$_8$ exhibits spin wave excitation which is quantitatively described by the nearest neighbor anisotropic Heisenberg model based on the $120^\circ$ spin structure.
In contrast, Li$_2$ScMo$_3$O$_8$ undergoes short-range magnetic order below 4~K with quantum-spin-liquid-like magnetic fluctuations down to the base temperature.
Origin of the different ground states is discussed in terms of anisotropies of crystal structures and magnetic interactions.
\end{abstract}

\maketitle

\section{Introduction}
When quantum spins are aligned on geometrically frustrated lattices, unusual ground state eventually emerges among energetically competed states~\cite{QSL1,QSL2,Kagome1}.
Two-dimensional (2D) spin-1/2 triangular lattice Heisenberg antiferromagnet (TLHAF) is a prototypical system of geometrically frustrated magnets.
Theoretically, the ground states of 2D TLHAF with both quantum and classical spins are known to be so-called $120^\circ$ long-range order~\cite{Classical1,Triangular1,Triangular2,Triangular3}.
When perturbations such as the second nearest-neighbor interaction~\cite{TQSL1}, ring exchange interaction~\cite{TQSL2}, spatially anisotropic interactions~\cite{TQSL3}, and randomness of the strength of the nearest-neighbor interaction~\cite{TQSL4} are set in, the system undergoes a quantum spin liquid (QSL) ground state where the system does not show static long-range magnetic order but shows long-range entanglement and fractional excitations~\cite{QSL1,QSL2}.
Extensive experimental studies have also been conducted on spin-1/2 TLHAFs; the $120^\circ$ long-range magnetic order is reported in Ba$_3$CoSb$_2$O$_9$~\cite{BCoSO1,BCoSO2,BCoSO3} whereas QSL state is proposed for the ground states of $\kappa$-(BEDT-TTF)$_2$Cu$_2$(CN)$_3$~\cite{BEDTTTF0,BEDTTTF1}, EtMe$_3$Sb[Pd(dmit)$_2$]$_2$~\cite{EtMe0,EtMe1}, YbMgGaO$_4$~\cite{YMGO1,YMGO2,YMGO3,YMGO4} and 1T-TaS$_2$~\cite{TaS21}.
Furthermore, spin-1 TLHAF Ba$_3$NiSb$_2$O$_9$ also shows QSL behaviors~\cite{BNSO1,BNSO2,BNSO3}.
QSL with spinon Fermi surface~\cite{SFS0,SFS1} was proposed and succeeded in understanding the QSL behaviors in such compounds~\cite{YMGO2,YMGO3,BNSO3}.
However, experimental realization of the QSL ground state in spin-1/2 TLHAF systems is still limited and remains an intriguing pursuit.

\begin{figure*}[ht]
\centering
\includegraphics[width=\linewidth]{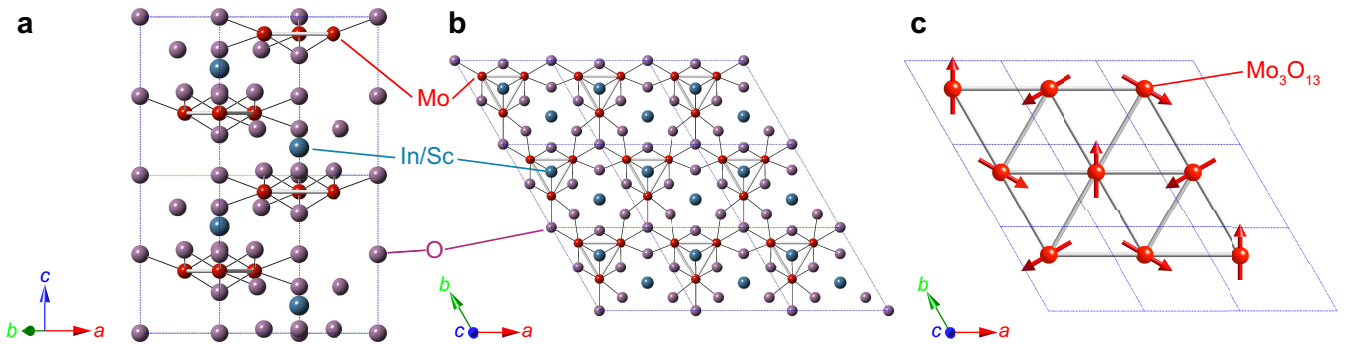}
\caption{
(a, b) Crystal structure of Li$_2$$A$Mo$_3$O$_8$.
Li site is not depicted.
(c) $120^\circ$ spin structure on the Mo$_3$O$_{13}$-based triangular lattice in Li$_2$InMo$_3$O$_8$.
Dashed lines in each panel represent the chemical unit cell.
}
\label{Fig:Structure}
\end{figure*}

Recently, cluster magnet LiZn$_2$Mo$_3$O$_8$ has attracted considerable research interest as spin-1/2 TLHAF~\cite{LZMO1}.
Seven $4d$ electrons in a Mo$_3$O$_{13}$ cluster occupy their orbitals, resulting in one unpaired electron.
Unpaired electron with spin $S=1/2$ remains in the total symmetry of the Mo$_3$O$_{13}$ cluster ($A_1$ irreducible representation) with equal contributions from all three Mo atoms, and network of the magnetic clusters forms a uniform triangular lattice in LiZn$_2$Mo$_3$O$_8$.
The dominant magnetic interaction between spin-1/2 Mo$_3$O$_{13}$ clusters is antiferromagnetic~\cite{LZMO1}, yielding geometrical frustration.
LiZn$_2$Mo$_3$O$_8$ is therefore an ideal 2D spin-1/2 TLHAF system.
Magnetic susceptibility and heat capacity measurements suggested that 2/3 of $S=1/2$ spins are quenched below 96~K, and condensed valence bond state (VBS) where resonance valence-bond states~\cite{RVB1,RVB2} coexist with remnant paramagnetic spins is proposed for the possible ground state~\cite{LZMO1,LZMO2}.
Gapless spin excitations were reported by electron spin resonance~\cite{LZMO2}, $^7$Li nuclear magnetic resonance (NMR)~\cite{LZMO2}, muon spin rotation ($\mu$SR)~\cite{LZMO2}, and inelastic neutron scattering (INS)~\cite{LZMO3} measurements.
Emergent honeycomb lattice is theoretically proposed for the origin of the condensed VBS~\cite{LZMO4}.
Recently, a 1/6-filled extended Hubbard model in an anisotropic kagome lattice is also proposed to account for the low temperature phase of LiZn$_2$Mo$_3$O$_8$~\cite{ChenTheory}.
However, intersite disorder between Li$^+$ and Zn$^{2+}$ ions is reported~\cite{LZMO1,LZMO2}, which may affect on the intrinsic magnetism in LiZn$_2$Mo$_3$O$_8$.

New molecular based triangular lattice systems Li$_2$$A$Mo$_3$O$_8$ where $A$ = In or Sc are of particular interest in this context~\cite{LIMO1,LAMO1}.
Li$_2$$A$Mo$_3$O$_8$ crystallizes in a hexagonal structure $P6_3mc$, and no intersite disorder between Li$^+$ and $A^{3+}$ sites exists (see Supplementary Information).
As in LiZn$_2$Mo$_3$O$_8$, spin-1/2 carrying Mo$_3$O$_{13}$ clusters are arranged on the structurally perfect triangular lattice separated by nonmagnetic Li and $A$ layers in both compounds as shown in Figs.~\ref{Fig:Structure}(a) and \ref{Fig:Structure}(b).
Susceptibility measurements of both compounds report that the dominant magnetic interactions are antiferromagnetic and the effective moments are $1.61\mu_\text{B}$ (In) and $1.65\mu_\text{B}$ (Sc), which are close to $p_\text{eff}=1.73\mu_\text{B}$ the ideal value for spin $S=1/2$.
Spin-1/2 TLHAF is therefore realized in Li$_2$$A$Mo$_3$O$_8$, whose ground states are however different from each other.
In Li$_2$InMo$_3$O$_8$, long-range magnetic order develops below $T_\text{N}=12$~K with Curie-Weiss temperature of $\Theta_\text{CW}=-242$~K, and $^7$Li NMR study suggests that the magnetic structure is the $120^\circ$ structure as described in Fig.~\ref{Fig:Structure}(c).
On the other hand, isostructural Li$_2$ScMo$_3$O$_8$ shows no long-range magnetic order down to 0.5~K in spite of large Weiss temperature of $\Theta_\text{CW}=-127$~K.
Instead, both magnetic susceptibility and heat capacity measurements indicate the development of short-range magnetic order below 10~K.
Spin glass state is ruled out as the ground state of Li$_2$ScMo$_3$O$_8$ since the magnetic susceptibility shows no splitting between zero-field-cooling and field-cooling processes~\cite{LAMO1}.
Low-temperature heat capacity measurements in Li$_2$ScMo$_3$O$_8$ shows sizable $T$-linear term $\gamma_\text{mag}=35.7$~mJ/mol$\cdot$K$^2$, which is similar to those of QSL candidates $\kappa$-(BEDT-TTF)$_2$Cu$_2$(CN)$_3$~\cite{BEDTTTF1}, EtMe$_3$Sb[Pd(dmit)$_2$]$_2$~\cite{EtMe1}, and Ba$_3$CuSb$_2$O$_9$~\cite{BCuSO1}.
Furthermore, different magnetic entropies between Li$_2$ScMo$_3$O$_8$ and LiZn$_2$Mo$_3$O$_8$ suggests that the ground state in Li$_2$ScMo$_3$O$_8$ is QSL rather than condensed VBS.
Because of easy access to two different ground states of spin-1/2 TLHAF, Li$_2$$A$Mo$_3$O$_8$ is an intriguing system to investigate 2D spin-1/2 TLHAF.
However, lack of microscopic measurements prevents us from fully understanding the ground states and dynamics of Li$_2$$A$Mo$_3$O$_8$.
In this paper, we investigate quantum magnetisms of polycrystalline Li$_2$InMo$_3$O$_8$ and Li$_2$ScMo$_3$O$_8$ by combination of $\mu$SR and time-of-flight (TOF) neutron scattering techniques.

\section{Experimental details}
The preparation of polycrystalline Li$_2$InMo$_3$0$_8$ (Li$_2$ScMo$_3$O$_8$) was carried out by two steps~\cite{LAMO1}.
First, to synthesize a precursor Li$_2$MoO$_4$, a mixture with a ratio of MoO$_3$ : Li$_2$CO$_3$ = 1:1 was ground, placed in an alumina crucible, and heated at 873~K for 24~hours in air; we repeated this step for three times.
Then, a mixture having a ratio of In$_2$O$_3$ (Sc$_2$O$_3$) : Li$_2$MoO$_4$ : MoO$_3$ : Mo = 0.5 : 1 : 0.84 : 1.16 was ground, pressed into a pellet, sealed in an evacuated quartz tube, heated at 923~K for 12 hours, and heated at 1198~K (1173~K) for 24 hours; we repeated this step for two times.
Magnetization measurements were performed using a commercial superconducting quantum interference device (SQUID) magnetometer (Quantum Design Magnetic Property Measurement System, MPMS).

ZF- and LF-$\mu$SR experiments were performed using the Advanced Research Targeted Experimental Muon Instrument at the S1 line spectrometer (ARTEMIS)~\cite{Muon1} with a conventional $^4$He flow cryostat and the D1 spectrometer~\cite{Muon1} with a $^3$He-$^4$He dilution refrigerator installed at Materials and Life Science Experimental Facility (MLF), Japan Proton Accelerator Research Complex (J-PARC).
We used the VASP software~\cite{VASP} for DFT calculation and the DipElec program~\cite{DipElec} to calculate the local magnetic fields in Li$_2$ScMo$_3$O$_8$.

TOF neutron scattering measurements were performed using the Fermi chopper spectrometer 4SEASONS at MLF, J-PARC~\cite{SIKI1}.
Frequencies of the Fermi chopper were 350 and 250~Hz for the In and Sc systems, resulting in the combinations of incident neutron energies of 11.9, 15.8, 22.0, and 32.7~meV, and 7.5, 10.3, 15.0, and 23.9~meV~\cite{SIKI2}, respectively.
A standard top-loading cryostat at 4SEASONS was used for the measurements on Li$_2$InMo$_3$O$_8$, whereas a $^4$He refrigerator and a $^3$He cryostat were used for Li$_2$ScMo$_3$O$_8$.
Empty can was measured at corresponding temperatures, and then subtracted from raw data of Li$_2$ScMo$_3$O$_8$.
TOF data were visualized by software suite Utsusemi~\cite{SIKI3}.
Neutron scattering intensities are converted to the absolute unit using the incoherent scattering of each sample~\cite{Abso} after correction of the neutron absorption effect.
Squared magnetic form factor of the Mo$_3$O$_{13}$ cluster~\cite{LZMO3} and $\hbar\omega$-dependent energy resolution at 4SEASONS~\cite{SIKI4} were included in the LSW calculations for Li$_2$InMo$_3$O$_8$.

\section{Results and Discussion}
\begin{figure*}[ht]
\centering
\includegraphics[width=\linewidth]{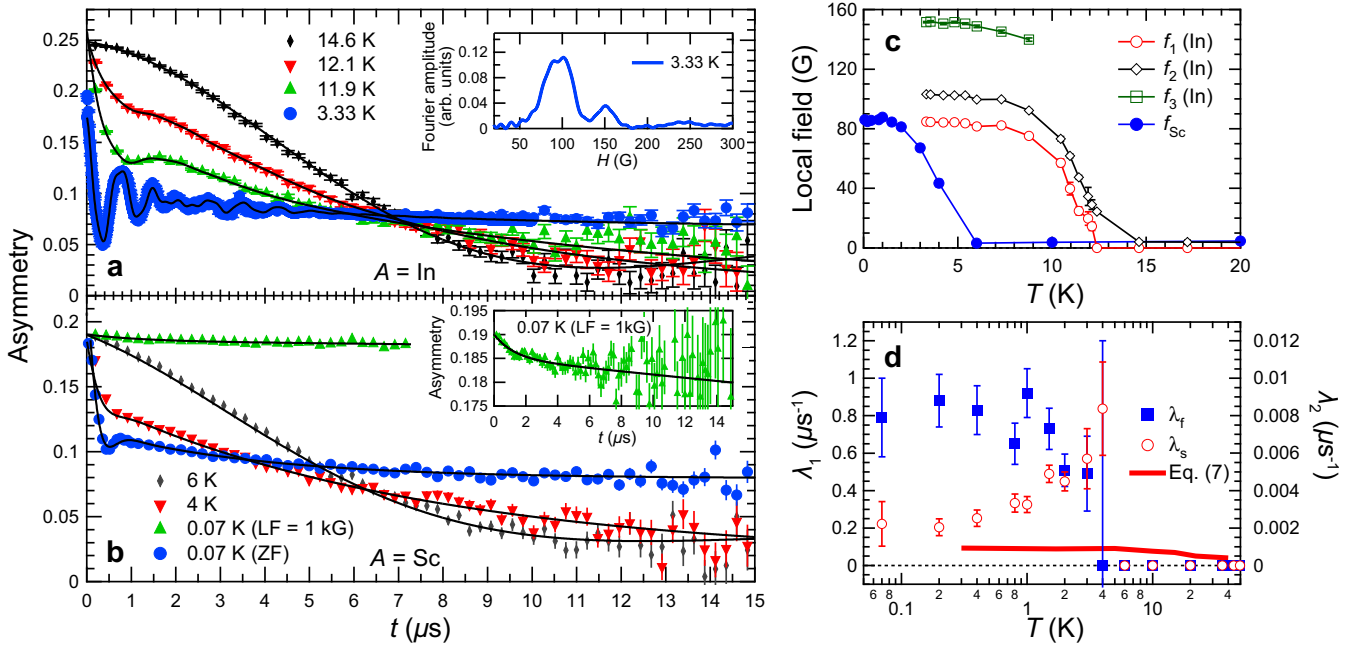}
\caption{
ZF- and LF-$\mu$SR results on Li$_2$$A$Mo$_3$O$_8$.
$\mu$SR time spectra of (a) Li$_2$InMo$_3$O$_8$ and (b) Li$_2$ScMo$_3$O$_8$.
Solid lines in each panel represent the fitting curves (see the main text).
A Fourier transform of the $\mu$SR time spectrum of Li$_2$InMo$_3$O$_8$ at 3.33~K (real amplitude) is plotted in the inset of (a).
The inset of (b) shows a magnified view of the spectrum of Li$_2$ScMo$_3$O$_8$ at 0.07~K under $H_\text{LF}=1$~kG.
(c) Temperature dependences of local fields in Li$_2$InMo$_3$O$_8$ and Li$_2$ScMo$_3$O$_8$.
(d) Temperature dependences of muon relaxation rates $\lambda_\text{f}$ (left scale) and $\lambda_\text{s}$ (right scale) of Li$_2$ScMo$_3$O$_8$ under longitudinal field.
Solid line is calculated $\lambda$ (right scale) using Eq.~(\ref{Eq4}).
}
\label{Fig:muSR}
\end{figure*}

Zero field- (ZF-) $\mu$SR time spectra of Li$_2$InMo$_3$O$_8$ at several temperatures are shown in Fig.~\ref{Fig:muSR}(a).
The spectra show a damping at around 12~K, and spectral oscillations appear at lower temperatures.
It is a direct evidence of the long-range magnetic order as reported in the earlier studies~\cite{LIMO1,LAMO1,LAMO3,LAMO4}.
Fourier transform of the spectrum at 3.33~K [see the inset of Fig.~\ref{Fig:muSR}(a)] suggests that at least three different local fields are found in Li$_2$InMo$_3$O$_8$, which is probably due to crystallographically inequivalent muon stopping sites indicated by our density functional theory (DFT) calculation (see Supplementary Fig.~S3 for Li$_2$ScMo$_3$O$_8$).
The ZF-$\mu$SR spectra of Li$_2$InMo$_3$O$_8$ are fitted by three cosine functions with transverse and longitudinal relaxations
\begin{eqnarray}
A_\text{ZF}(t)&=&\sum_{n=1}^3A_n[\frac{2}{3}\text{cos}\left(2\pi f_nt+\phi\right)\text{exp}\left(-\lambda_tt\right)\nonumber\\
&+&\frac{1}{3}\text{exp}\left(-\lambda_lt\right)]+A_\text{BG}\label{Eq_ZFIn}
\end{eqnarray}
where $A_n$ and $A_\text{BG}$ are the positron decay asymmetries of each oscillation ($n=1\sim3$) and background (mainly from a silver backing plate) components, $f_n$ is the precession frequency, $\phi$ is the initial phase, and $\lambda_t$ ($\lambda_l$) is the transverse (longitudinal) relaxation rate.
Fitting result at each temperature is shown in Fig.~\ref{Fig:muSR}(a).
Local magnetic fields of 84.9(3), 103.1(2), and 151.5(5)~G are extracted at 3.33~K, and these values are comparable in magnitude of local fields that are observed in spin-1/2 magnets~\cite{Cup1}.
Figure~\ref{Fig:muSR}(c) shows temperature dependences of $f_1$, $f_2$, and $f_3$, representing that long-range magnetic order evolves in Li$_2$InMo$_3$O$_8$ below $T_\text{N}=12$~K with the critical exponents $\beta\sim0.33$.

In the meanwhile, ZF-$\mu$SR time spectrum of Li$_2$ScMo$_3$O$_8$ at 0.07~K shows a highly damped oscillation with a pronounced reduction of the 1/3 tail as described in Fig.~\ref{Fig:muSR}(b).
To see the temperature evolution of the local fields in Li$_2$ScMo$_3$O$_8$, the ZF-$\mu$SR spectrum  are fitted by combination of transverse and longitudinal relaxations
\begin{eqnarray}
A_\text{ZF}(t)&=&A_1\text{cos}\left(2\pi f_\text{Sc}t+\phi\right)\text{exp}\left(-\lambda_tt\right)\nonumber\\
&+&A_2\text{exp}\left(-\lambda_lt\right)+A_\text{BG}.\label{Eq_ZFSc}
\end{eqnarray}
The fitting result at each temperature is plotted in Fig.~\ref{Fig:muSR}(b), and temperature dependence of the local field $f_\text{Sc}$ is also plotted in Fig.~\ref{Fig:muSR}(c).
One can clearly see the temperature evolution of $f_\text{Sc}$ below 4~K with the critical exponent $\beta\sim0.28$ which is similar to those of Li$_2$InMo$_3$O$_8$.
Therefore, magnetic nature of these compounds are essentially the same, but it should be noted that the ground state of Li$_2$ScMo$_3$O$_8$ is short-range magnetic order by considering the strong damping of the oscillation below 4~K.
The anomaly at $4$~K was also found in the temperature derivative of the magnetic susceptibility~\cite{LAMO1}.
Although the short-range magnetic order develops in Li$_2$ScMo$_3$O$_8$ below 4~K, the spectrum shows a moderate tail over a long period of time, suggesting that spin fluctuation survives even at 0.07~K.
To explicitly distinguish the spin fluctuation of the Mo$_3$O$_{13}$ cluster, we performed longitudinal field- (LF-) $\mu$SR measurements on Li$_2$ScMo$_3$O$_8$ under longitudinal magnetic field ($H_\text{LF}$) of 1~kG.
Figure~\ref{Fig:muSR}(b) and its inset display a LF-$\mu$SR time spectrum measured at 0.07~K.
$H_\text{LF}=1$~kG seems to be sufficient to quench (decouple) muon spin relaxations by both nuclear dipoles and the short-range ordered state.
The characteristic LF-$\mu$SR spectrum of Li$_2$ScMo$_3$O$_8$ at 0.07~K was fitted by the following equation
\begin{equation}
A_\text{LF}(t)=A_\text{f}\text{exp}(-\lambda_\text{f}t)+A_\text{s}\text{exp}(-\lambda_\text{s}t)+A_\text{BG}
\end{equation}
where $A_\text{f}$ and $A_\text{s}$ are asymmetries of fast ($\lambda_\text{f}$) and slow ($\lambda_\text{s}$) relaxation components, respectively ($A_\text{f}+A_\text{s}=0.16$), and $A_\text{BG}$ is the background asymmetry ($A_\text{BG}=0.03$).
The fitting results are described by the solid lines in Fig.~\ref{Fig:muSR}(b) and its inset.
We also fit LF-$\mu$SR time spectra under $H_\text{LF}=1$~kG at several temperatures, and obtained temperature dependences of $\lambda_\text{f}$ and $\lambda_\text{s}$ are plotted in Fig.~\ref{Fig:muSR}(d).
$\lambda_\text{f}$ shows a rapid relaxation with relative signal amplitude of $\sim3$\%.
It mainly corresponds to the remnant signal from the short-range ordered state since $\lambda_\text{f}$ exhibits a steep increase at 4~K as temperature goes down.
On the other hand, $\lambda_\text{s}$ shows a slow relaxation with two orders of magnitude less than $\lambda_\text{f}$, which is related to the intrinsic spin fluctuation of the Mo$_3$O$_{13}$ cluster.
Remarkably, temperature dependence of $\lambda_\text{s}$ shows a temperature-independent plateau below 1~K and converges into the finite value of $\sim0.002$~$\mu$s$^{-1}$ which is very close to that of triangular lattice QSL 1T-TaS$_2$ ($\lambda=0.0023$~$\mu$s$^{-1}$ at 0.07~K)~\cite{TaS21}.
Indeed, such low-temperature plateau behaviors of muon relaxation rate is common feature in the TLHAF QSL candidates~\cite{YMGO2,BNSO2}, which will be discussed again.
To obtain complementary information to our $\mu$SR results on Li$_2$$A$Mo$_3$O$_8$, TOF neutron scattering measurements were also conducted.

\begin{figure*}[ht]
\centering
\includegraphics[width=\linewidth]{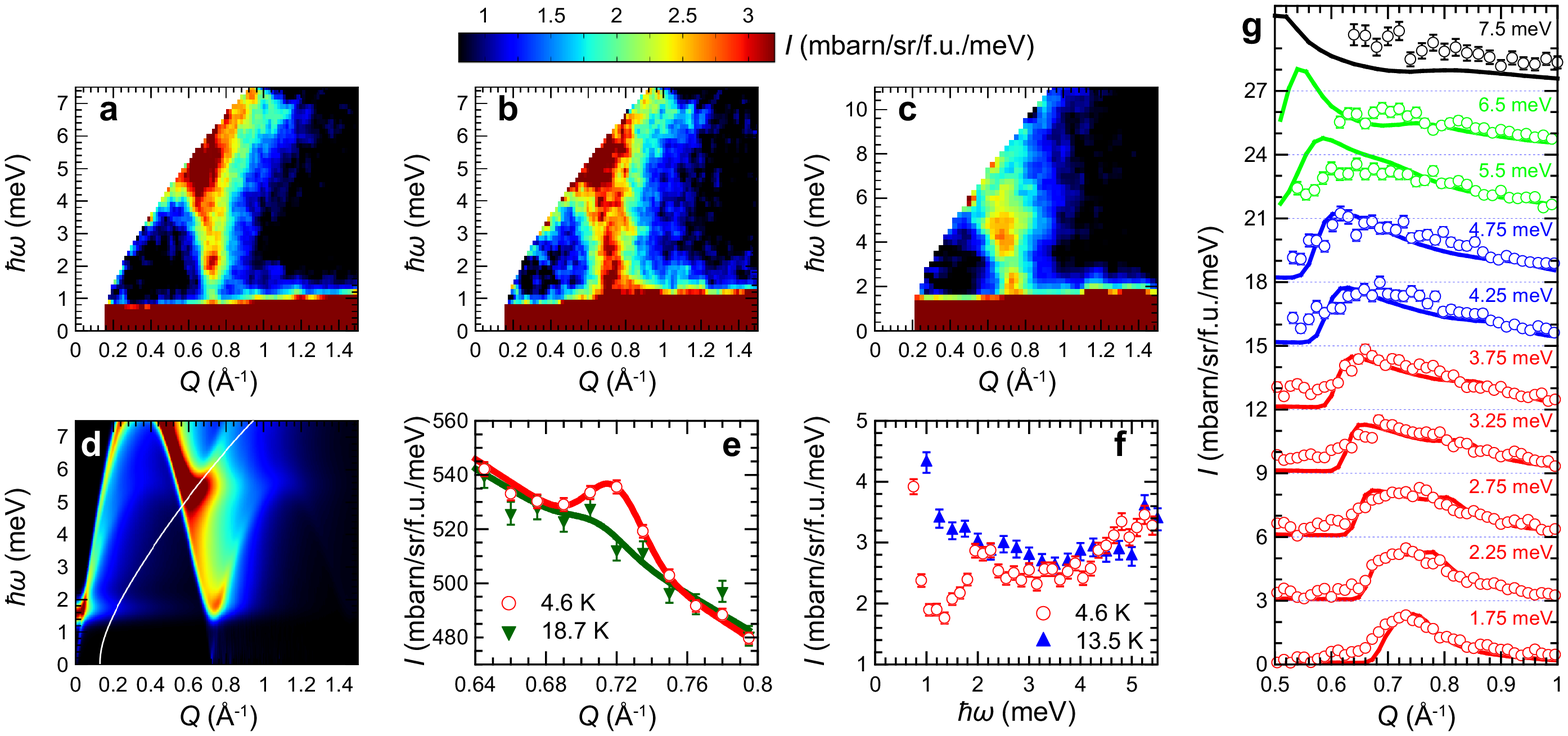}
\caption{
TOF neutron scattering results on Li$_2$InMo$_3$O$_8$.
Low-energy inelastic neutron scattering intensity maps at (a) 4.6~K and (b) 13.5~K measured with $E_\text{i}=11.9$~meV.
(c) High-energy inelastic neutron scattering intensity map at 4.6~K measured with $E_\text{i}=22.0$~meV.
(d) Calculated inelastic neutron scattering intensity map at 4.6~K using the optimum parameters as described in the main text.
Calculated energy resolution for $E_\text{i}=11.9$~meV was convoluted.
(e) Elastic neutron scattering intensities at 4.6 and 18.7~K with energy window of $[-0.15,0.15]$~meV.
Solid lines represent the fitting result using the Gaussian function with linear background.
(f) Energy spectra integrated over $Q=[0.69,0.77]$~$\text{\AA}^{-1}$ at 4.6 and 13.5~K measured with $E_\text{i}=11.9$~meV.
(g) $Q$ dependences of the neutron scattering intensities at several energy windows.
Each energy window was [1.5, 2.0], [2.0, 2.5], [2.5, 3.0], [3.0, 3.5], and [3.5, 4.0]~meV with $E_\text{i}=11.9$~meV (red), [4.0, 4.5] and [4.5, 5.0]~meV with $E_\text{i}=15.8$~meV (blue), [5.0, 6.0] and [6.0, 7.0 ]~meV with $E_\text{i}=22.0$~meV (green), and [7.0, 8.0]~meV with $E_\text{i}=32.7$~meV (black), respectively.
Constant background was subtracted from each $Q$ dependence.
Solid lines are calculated results using the optimum parameters in Eq.~(\ref{parameter}).
}
\label{Fig:Li2InMo3O8}
\end{figure*}

Elastic neutron scattering spectra of Li$_2$InMo$_3$O$_8$ below and above $T_\text{N}$ are shown in Fig.~\ref{Fig:Li2InMo3O8}(e).
A magnetic Bragg peak appears at momentum transfer $Q=0.719(1)$~$\text{\AA}^{-1}$ below $T_\text{N}$.
The $Q$ position corresponds to $(1/3, 1/3, 0)$, indicating the $120^\circ$ magnetic structure consistent with the previous $^7$Li-NMR measurements~\cite{LAMO1}.
By comparing the intensity of the magnetic peak with those of nuclear Bragg peaks, the ordered moment at 4.6~K is estimated to be $0.51(3)\mu_\text{B}$.
Theoretically, the magnetic moment is reduced by about 59\% for the spin-1/2 TLHAF~\cite{Triangular2}, which is close to the observed ordered moment (reduced by 49\% assuming $g=2$).
The reduced moment originates in a combination of geometrical frustration and quantum fluctuation.
Neutron scattering intensity ($I$) map from Li$_2$InMo$_3$O$_8$ as a function of $Q$ and energy transfer ($\hbar\omega$) at 4.5~K ($<T_\text{N}$) is shown in Fig.~\ref{Fig:Li2InMo3O8}(a).
Dispersive excitation centered at the magnetic zone center $(1/3, 1/3, 0)$ was observed.
Because of the $Q$ position, the excitation is assigned to be the spin wave excitation in the long-range magnetic ordered state.
Energy spectrum at the magnetic zone center exhibits a substantial peak at $\hbar\omega=2.08(3)$~meV as shown in Fig.~\ref{Fig:Li2InMo3O8}(f).
This result claims that one branch (or some branches) of the spin wave excitation has spin gap at the magnetic zone center due to the magnetic anisotropy.
On the other hand, magnetic signals at the magnetic zone center become quasielastic above $T_\text{N}$ as shown in Figs.~\ref{Fig:Li2InMo3O8}(b) and \ref{Fig:Li2InMo3O8}(f).
Therefore, the gap-like excitation is a characteristic feature of the long-range magnetic ordered state.
To observe the whole structure of the spin wave excitation at 4.5~K, $I(Q,\hbar\omega)$ map using higher $E_\text{i}$ is presented in Fig.~\ref{Fig:Li2InMo3O8}(c).
The spin wave excitation survives up to $\sim9$~meV.
$Q$ dependences of the spin wave intensities at various $\hbar\omega$s are plotted in Fig.~\ref{Fig:Li2InMo3O8}(g).
The spectra are asymmetric at $\hbar\omega>3.0$~meV, and the peak shifts to lower $Q$ at higher $\hbar\omega$.
This result suggests that the squared magnetic form factor ($|F(Q)|^2$) of the Mo$_3$O$_{13}$ cluster decreases quickly and is negligible at high $Q$, representing the unpaired electron with equal contributions from all three Mo atoms in Li$_2$InMo$_3$O$_8$.

For quantitative analysis on the spin wave excitation in Li$_2$InMo$_3$O$_8$, semi-classical linear spin wave (LSW) analysis was performed considering the $120^\circ$ spin structure on the spin-1/2 2D Mo$_3$O$_{13}$-based triangular lattice [Fig.~\ref{Fig:Structure}(c)].
The gap-like excitation at the magnetic zone center in the long-range magnetic ordered state is also observed in the other spin-1/2 triangular lattice system Ba$_3$CoSb$_2$O$_9$~\cite{BCoSO2}, and the peak energy ($E_0$) roughly scales with $T_\text{N}$ in these compounds: $E_0=0.65$~meV and $T_\text{N}=3.8$~K in Ba$_3$CoSb$_2$O$_9$~\cite{BCoSO2} whereas $E_0=$2.08~meV and $T_\text{N}=12$~K in Li$_2$InMo$_3$O$_8$.
This suggests that the origin of the gap-like excitation in Li$_2$InMo$_3$O$_8$ is the same as that in Ba$_3$CoSb$_2$O$_9$~\cite{BCoSO2}.
Therefore, as in Ba$_3$CoSb$_2$O$_9$~\cite{BCoSO1,BCoSO2,BCoSO3}, the nearest-neighbor anisotropic exchange interaction was considered as the model Hamiltonian for Li$_2$InMo$_3$O$_8$
\begin{equation}
\mathcal{H}=\alpha J\sum_{i,j}(S_i^xS_j^x+S_i^yS_j^y+\delta S_i^zS_j^z)
\end{equation}
where $\alpha$, $J$, and $\delta$ represent the renormalization factor, the nearest neighbor exchange coupling constant, and the anisotropic factor.
$J$ was fixed to 112~K determined by the magnetic susceptibility measurement~\cite{LAMO1}.
By fitting the calculated powder-averaged $Q$ dependences to the experimental results at different $\hbar\omega$s ($2\sim7.5$~meV) simultaneously, optimum parameters were yielded
\begin{eqnarray}
\alpha&=&0.56(1), \nonumber\\
\delta&=&0.975(1).\label{parameter}
\end{eqnarray}
Fitting results together with the experimental results are shown in Fig.~\ref{Fig:Li2InMo3O8}(g), and calculated LSW $I(Q,\hbar\omega)$ map is also shown in Fig.~\ref{Fig:Li2InMo3O8}(d).
Satisfactory agreements with calculation and experiment were confirmed.
Obtained $\alpha$ is smaller than 1, indicating a negative quantum renormalization effect theoretically proposed for 2D spin-1/2 TLHAF~\cite{Reno1,Reno2,Reno3}.
Similar negative quantum renormalization effect ($\alpha\sim0.65$) was also reported in Ba$_3$CoSb$_2$O$_9$~\cite{BCoSO3}.
Therefore, observed magnetic excitations of Li$_2$InMo$_3$O$_8$ in the accessible $(Q,\hbar\omega)$ region are well understood by the semi-classical LSW theory assuming the $120^\circ$ magnetic structure on the spin-1/2 Mo$_3$O$_{13}$ triangular lattice.

\begin{figure*}[ht]
\centering
\includegraphics[width=\linewidth]{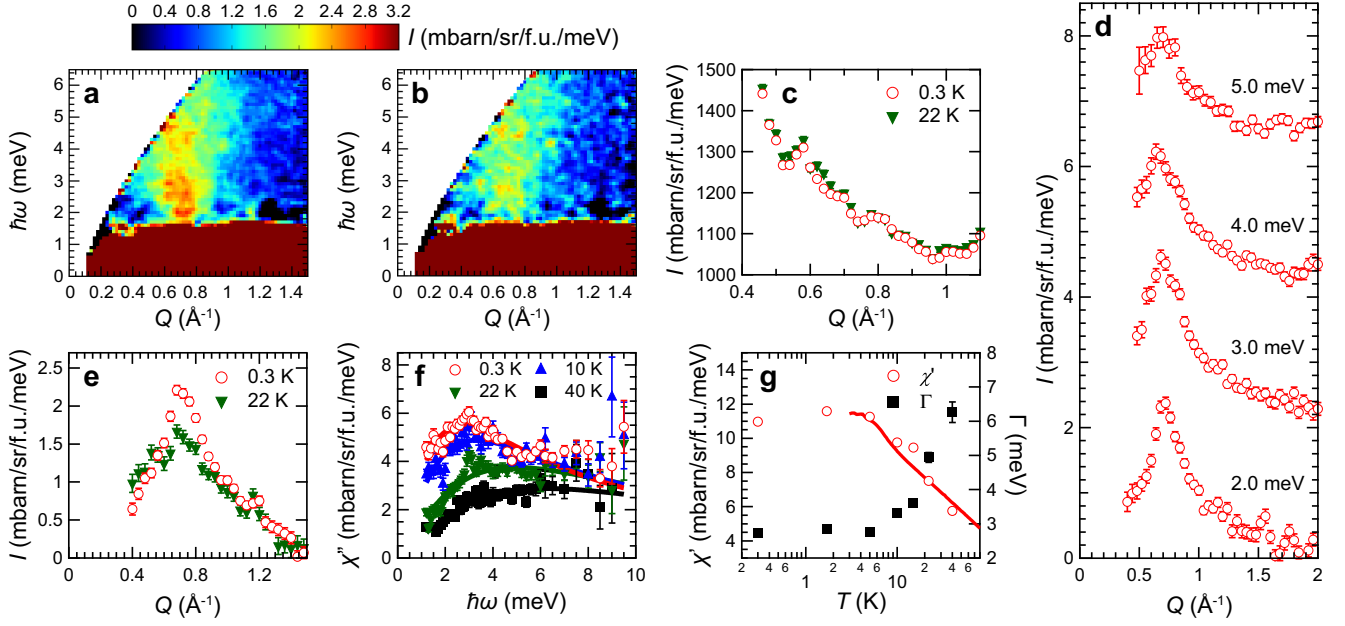}
\caption{
TOF neutron scattering results on Li$_2$ScMo$_3$O$_8$.
Inelastic neutron scattering intensity maps with $E_\text{i}=10.3$~meV measured at (a) 0.3~K and (b) 22~K.
(c) Elastic neutron scattering intensities with energy window of $[-0.075,0.075]$~meV at 0.3 and 22~K with $E_\text{i}=10.3$~meV.
(d) $Q$ dependences of the neutron scattering intensities at several energy windows at 1.7~K.
Each energy window was $[1.5, 2.5]$~meV with $E_\text{i}=7.5$~meV, $[2.5, 3.5]$~meV with $E_\text{i}=10.3$~meV, $[3.5, 4.5]$~meV with $E_\text{i}=15.0$~meV, and [4.5, 5.5]~meV with $E_\text{i}=23.9$~meV, respectively.
(e) $Q$ dependences of the neutron scattering intensities with $[1.5, 2.5]$~meV at 0.3 and 22~K using $E_\text{i}=$7.5~meV.
(f) Dynamical spin susceptibilities at $Q=[0.6, 0.8]$~$\text{\AA}^{-1}$ for 0.3, 10, 22, and 40~K.
Solid lines are the fitting results by the quasielastic Lorentzian as described in the main text.
(g) Temperature dependences of static spin susceptibility $\chi'$ (left scale) and relaxation rate $\Gamma$ (right scale) obtained by the fitting in panel (f).
Solid line is a scaled bulk magnetic susceptibility ($\chi_\text{bulk}=M/H$) measured with $H=1$~T.
}
\label{Fig:Li2ScMo3O8}
\end{figure*}

In contrast to Li$_2$InMo$_3$O$_8$, no magnetic Bragg peak evolves in the elastic channel down to 0.3~K in Li$_2$ScMo$_3$O$_8$ as plotted in Fig.~\ref{Fig:Li2ScMo3O8}(c), in agreement with our $\mu$SR results.
On the other hand, diffuse scattering expected for the short-range order is not observed in our neutron measurements.
Strong incoherent scattering may smear out such magnetic diffuse scattering in Li$_2$ScMo$_3$O$_8$.
Figure~\ref{Fig:Li2ScMo3O8}(a) depicts $I(Q,\hbar\omega)$ map at 0.3~K.
Clear diffuse scattering was observed in the inelastic channel.
Although both magnetic excitations in Li$_2$InMo$_3$O$_8$ and Li$_2$ScMo$_3$O$_8$ are centered at $Q\sim0.7$~$\text{\AA}^{-1}$ [Figs.~\ref{Fig:Li2InMo3O8}(a) and \ref{Fig:Li2ScMo3O8}(a)], the overall structures are different, representing their different ground states.
In Li$_2$ScMo$_3$O$_8$, steep continuum excitation was observed.
The $Q$ dependences of the magnetic excitations are invariant in the different energy windows as shown in Fig.~\ref{Fig:Li2ScMo3O8}(d).
Steep continuum excitation, or spinon continuum, is the common feature of the magnetic excitations in the QSL candidates~\cite{Kagome1,YMGO3,YMGO4,BNSO3,Kape1}.
$I(Q,\hbar\omega)$ map at high temperature (22~K) is also shown in Fig.~\ref{Fig:Li2ScMo3O8}(b).
Although overall magnetic fluctuation at 22~K is similar to that at 0.3~K, there are some differences.
Scattering intensity decreases at 22~K.
In addition, as shown in Fig.~\ref{Fig:Li2ScMo3O8}(e), spectrum weight of the $Q$ dependence at 2~meV slightly shifts to $Q=0$ at high temperature, which is also observed in other QSL candidates~\cite{YMGO4,BNSO3,Kape1}.

To investigate in more detail the characteristic energy (or time) scale of the steep continuum in Li$_2$ScMo$_3$O$_8$, the dynamical spin susceptibilities $\chi"(\hbar\omega)=[1-\text{exp}(-\hbar\omega/k_\text{B}T)]/|F(Q)|^2I(\hbar\omega)$ at $Q=[0.6, 0.8]$~$\text{\AA}^{-1}$ where the magnetic signal is maximal are plotted for different temperatures in Fig.~\ref{Fig:Li2ScMo3O8}(f).
The spectra are well fitted by the quasielastic Lorentzian $\chi"(\hbar\omega)=\chi'\hbar\omega\Gamma/[(\hbar\omega)^2+\Gamma^2]$ where $\chi'$ is the static susceptibility and $\Gamma$ the spin relaxation rate [or peak position of $\chi"(\hbar\omega)$].
The temperature dependences of the resulting parameters are shown in Fig.~\ref{Fig:Li2ScMo3O8}(g).
Upon decreasing temperature, $\Gamma$ decreases while $\chi'$ increases.
Contrary to the conventional long-range ordered magnets, no divergent behavior was observed in the temperature dependences of $\chi'$ and $\Gamma$.
It should be noted that $\chi'$ scales with bulk magnetic susceptibility $\chi_\text{bulk}$ over the temperature range of $3\le T\le40$~K [see solid line in Fig.~\ref{Fig:Li2ScMo3O8}(g)] and $\Gamma$ is also scaled by the muon relaxation rate $\lambda_\text{s}$ as discussed below.
These fittings also extract two important features of the steep continuum in Li$_2$ScMo$_3$O$_8$: (1) the magnetic excitation is gapless consistent with the heat capacity measurement~\cite{LAMO1} and (2) the dynamical spin susceptibility extends from the elastic channel up to at least 9.5~meV which is about $1.6J$ where $J$ ($=67$~K) is determined by the magnetic susceptibility measurement~\cite{LAMO1}.

Complementary analysis of $\mu$SR and INS results enables us to exclusively clarify the quantum fluctuations in Li$_2$ScMo$_3$O$_8$.
Muon spin relaxation rate $\lambda_\text{s}$ in Fig.~\ref{Fig:muSR}(d) is related to the spin relaxation rate of the magnetic fluctuation $\Gamma$ in Fig.~\ref{Fig:Li2ScMo3O8}(g) on the basis of following Redfield's formula~\cite{Redfield}
\begin{equation}
\lambda=\frac{2(\gamma_\mu\delta_\mu)^2\Gamma}{(\gamma_\mu H_\text{LF})^2+\Gamma^2}\label{Eq3}
\end{equation}
where $\gamma_\mu$ and $\delta_\mu$ are the gyromagnetic ratio of muon (=$2\pi\times135.54$~MHz/T) and average distribution of local magnetic fields at muon sites.
We performed electrostatic potential calculations using a point-charge model~\cite{DipElec} and estimated $\delta_\mu=204.8$~G for Li$_2$ScMo$_3$O$_8$ (see Supplementary Information).
Since $H_\text{LF}$ ($=1$~kG $=8.5\times10^8$~Hz) is much smaller than $\Gamma$ ($=2.7$~meV $=6.5\times10^{11}$~Hz at 0.3 K) in Li$_2$ScMo$_3$O$_8$, Eq.~(\ref{Eq3}) is reformulated as 
\begin{equation}
\lambda\sim\frac{2(\gamma_\mu\delta_\mu)^2}{\Gamma}.\label{Eq4}
\end{equation}
We plotted calculated temperature dependence of $\lambda$ using $\Gamma$ obtained by our INS measurements and compared with $\lambda_\text{s}$ obtained by our LF-$\mu$SR measurements [see solid line for calculation and circles for $\mu$SR results in Fig.~\ref{Fig:muSR}(d)].
Quantitative agreement can be seen; the anomaly around 4~K is artificial feature owing to $\lambda_\text{f}$.
Therefore, both $\mu$SR and INS measurements exhibit that quantum fluctuations persist at the lowest measured temperature.
As mentioned above, such low-temperature plateaus of the relaxation rates were widely observed in the triangular-lattice~\cite{YMGO2,BNSO2} and kagome-lattice~\cite{Kape1,Kagome2,Kagome3,Kagome4,Kagome5} QSL candidates.

To account for the QSL-like excitations in Li$_2$ScMo$_3$O$_8$, we now consider the spinon Fermi surface QSL model.
In Li$_2$ScMo$_3$O$_8$, no static long-range order was detected even down to 0.07~K [Figs.~\ref{Fig:muSR}(b) and \ref{Fig:Li2ScMo3O8}(c)].
Alternatively, gapless continuum in Li$_2$ScMo$_3$O$_8$ was observed at $Q=0.726(4)$~$\text{\AA}^{-1}$ corresponding to the $(1/3, 1/3, 0)$ position [Figs.~\ref{Fig:Li2ScMo3O8}(a), \ref{Fig:Li2ScMo3O8}(d), and \ref{Fig:Li2ScMo3O8}(f)].
Moreover, both $\lambda_\text{s}$ and $\Gamma$ exhibit temperature-independent plateaus at low temperature [Figs.~\ref{Fig:muSR}(d) and \ref{Fig:Li2ScMo3O8}(g)].
These features are well explained by QSL with spinon Fermi surface~\cite{SFS0,SFS1}.
As discussed in earlier works~\cite{YMGO2,YMGO3,BNSO3}, the spinon Fermi surface QSL model on the spin-1/2 TLHAF expects that (1) absence of static long-range magnetic order, (2) muon spin relaxation rate approach a finite value as temperature approaches zero, (3) magnetic excitation is gapless continuum, and (4) $\chi"(Q,\hbar\omega)$ shows the maximum intensity at the corner of the 2D Brillouin zone [e.g. $(1/3, 1/3, 0)$].
All observed features of the magnetic fluctuation in Li$_2$ScMo$_3$O$_8$ can be well described by the spinon Fermi surface QSL model.
Although the second peak of the spinon continuum in Ba$_3$NiSb$_2$O$_9$ was also observed at $(2/3, 2/3, 0)$~\cite{BNSO3}, the second peak in Li$_2$ScMo$_3$O$_8$ was not detected at $(2/3, 2/3, 0)$ corresponding to $Q=1.45$~$\text{\AA}^{-1}$ as shown in Fig.~\ref{Fig:Li2ScMo3O8}(d) because of the quick decay of the squared magnetic form factor of the Mo$_3$O$_{13}$ cluster~\cite{LZMO3}.
By performing complementary analysis on $\mu$SR and INS results, we conclude that Li$_2$ScMo$_3$O$_8$ undergoes the short-range magnetic order below 4~K with the QSL-like fluctuations which persist down to the lowest temperature.

We compare the Mo$_3$O$_{13}$-cluster-based triangular lattice antiferromagnets, Li$_2$$A$Mo$_3$O$_8$ and LiZn$_2$Mo$_3$O$_8$, in line with the recent theory by Chen $et$ $al$~\cite{ChenTheory}.
They proposed a 1/6-filled Hubbard model on an anisotropic kagome lattice with the nearest-neighbor electron hopping and repulsions~\cite{ChenTheory} to account for the magnetism in LiZn$_2$Mo$_3$O$_8$~\cite{LZMO1}.
Electron is fractionalized into charged boson and spin-carring spinons; plaquette charge order emerges as the charge ground state and the spin degree of freedom can be then described by $U(1)$ QSL with spinon Fermi surface, which can explain the unusual magnetic susceptibility in LiZn$_2$Mo$_3$O$_8$~\cite{LZMO1}.
For comparison with different compounds, they introduce a phenomenological parameter $\xi$ to characterize the anisotropy of the Mo kagome lattice: $\xi=d_\text{inter}/d_\text{intra}$ where $d_\text{intra}$ ($d_\text{inter}$) is the intracluster (intercluster) Mo-Mo bond length.
Large anisotropy $\xi$ tends to suppress charge fluctuations between clusters leading to the $120^\circ$ long-range magnetic order whereas small anisotropy $\xi$ corresponds to large charge fluctuation generating the $U(1)$ QSL with spinon Fermi surface.
Using the structural parameters summarized in Supplementary Information, we estimated $\xi$ as 1.271, 1.269, and 1.258 for Li$_2$InMo$_3$O$_8$, Li$_2$ScMo$_3$O$_8$, and LiZn$_2$Mo$_3$O$_8$~\cite{LZMO1}, respectively.
The phenomenological parameter $\xi$ explains the different ground states between the $120^\circ$ long-range magnetic order in Li$_2$InMo$_3$O$_8$ and the condensed VBS in LiZn$_2$Mo$_3$O$_8$.
However, the $\xi$ values of Li$_2$InMo$_3$O$_8$ and Li$_2$ScMo$_3$O$_8$ are very close to each other in spite of their different ground states.
Nevertheless, $^{115}$In and $^{45}$Sc NMR measurements on Li$_2$$A$Mo$_3$O$_8$ reported that charge fluctuation in Li$_2$ScMo$_3$O$_8$ is 2.6 times larger than that in Li$_2$InMo$_3$O$_8$~\cite{LAMO2}, and the difference between charge fluctuations can qualitatively explain the different ground states of Li$_2$InMo$_3$O$_8$ and Li$_2$ScMo$_3$O$_8$.
Therefore, the anisotropic parameter for the Mo kagome lattice, $\xi$, is too simplified to explain the different ground states in Li$_2$$A$Mo$_3$O$_8$, and more detailed parameter is required for Li$_2$$A$Mo$_3$O$_8$.

We also compare the magnetic excitations in Li$_2$InMo$_3$O$_8$ and Li$_2$ScMo$_3$O$_8$ to discuss the origin of the different ground states.
Although both magnetic excitations in Li$_2$InMo$_3$O$_8$ and Li$_2$ScMo$_3$O$_8$ center at $Q\sim0.72$~$\text{\AA}^{-1}$, low-energy magnetic excitations show opposite behaviors.
The magnetic excitation at the magnetic zone center in Li$_2$InMo$_3$O$_8$ clearly exhibits the peak at 2.08(3)~meV [Figs.~\ref{Fig:Li2InMo3O8}(a) and \ref{Fig:Li2InMo3O8}(f)].
Our LSW analysis suggests that the anisotropic exchange interaction is necessary to reproduce the peak.
Meanwhile, the gapless magnetic excitation in Li$_2$ScMo$_3$O$_8$ indicates that magnetic anisotropy is negligibly small in Li$_2$ScMo$_3$O$_8$ [Figs.~\ref{Fig:Li2ScMo3O8}(a) and \ref{Fig:Li2ScMo3O8}(f)].
Thus, the difference in the magnetic anisotropy is another possibility of the origin of the different ground states in Li$_2$$A$Mo$_3$O$_8$.
In fact, the gap-like excitation was observed in the long-range ordered state of Ba$_3$CoSb$_2$O$_9$~\cite{BCoSO2,BCoSO3}, whereas the gapless magnetic excitations in the QSL systems YbMgGaO$_4$~\cite{YMGO3,YMGO4} and Ba$_3$NiSb$_2$O$_9$~\cite{BNSO3}.
INS measurements on magnetic excitations in the substitution system Li$_2$(In$_{1-x}$Sc$_x$)Mo$_3$O$_8$~\cite{LAMO4} are effective to further elucidate the origin of different magnetic ground states, which is left for future work.

\section{Conclusion}
We performed a comprehensive study on the quantum magnetisms in the Mo$_3$O$_{13}$-cluster-based spin-1/2 triangular lattice antiferromagnets, Li$_2$InMo$_3$O$_8$ and Li$_2$ScMo$_3$O$_8$ by means of $\mu$SR and TOF neutron scattering techniques.
Spin wave excitation in Li$_2$InMo$_3$O$_8$ was well described by the nearest neighbor anisotropic Heisenberg model based on the $120^\circ$ spin structure.
Li$_2$ScMo$_3$O$_8$ exhibits the short-range magnetic order below 4~K with the QSL-like fluctuations which persist down to the lowest temperature.
The origin of the different magnetic ground states in Li$_2$$A$Mo$_3$O$_8$ is discussed in terms of anisotropies of crystal structures and magnetic interactions.

\section*{Acknowledgements}
We acknowledge Seiko Ohira-Kawamura for helpful discussion and Hua Li for DFT calculations.
We also thank technical supports from the MLF sample environment team.
Magnetic susceptibility measurements were performed at the CROSS user laboratories.
The synchrotron X-ray diffraction experiments were conducted at BL5S2 of Aichi Synchrotron Radiation Center, Aichi Science and Technology Foundation, Aichi, Japan (Proposal Nos. 201803046 and 201803047).
The $\mu$SR measurements at the S1 and D1 beamlines were conducted under the user program with proposal number 2017B0033.
The proposal numbers for the TOF neutron scattering experiments at 4SEASONS were 2016I0001, 2017A0004, 2017B0030, and 2018I0001.
The DFT calculations were supported by the Condensed Matter Research Center, Institute of Materials Structure Science, KEK, and the KEK Large Scale Simulation Program (Nos. 15/16-07 and 16/17-18).
This work was partially supported by JSPS KAKENHI Grant Numbers JP17K14349 and JP18K03529, and the Cooperative Research Program of ``Network Joint Research Center for Materials and Devices'' (20181072).

\end{document}